\documentclass[11pt]{article}
\usepackage[]{geometry} 
\usepackage{authblk}
\usepackage{amsmath, amssymb, enumerate, amsthm,amssymb,mathtools}
\usepackage{hyperref}
\usepackage{natbib}
\usepackage[normalem]{ulem}

\begin{document}

\title{Bayesian size-and-shape regression modelling}

\author[1,2]{{Antonio} {Di Noia}}
\affil[1]{Seminar for Statistics, Department of Mathematics, ETH Zurich}
\affil[2]{Faculty of Economics, Università della Svizzera italiana}
\author[3]{{Gianluca} {Mastrantonio}}
\affil[3]{Department of Mathematical Sciences, Polytechnic of Turin}
\author[4]{{Giovanna} {Jona Lasinio}}
\affil[4]{Department of Statistical Sciences, Sapienza University}

	\date{}

	\maketitle
	\let\thefootnote\relax\footnotetext{\emph{Email addresses:} antonio.dinoia(\sout{at})stat.math.ethz.ch (A. Di Noia), gianluca.mastrantonio(\sout{at})polito.it (G. Mastrantonio), giovanna.jonalasinio(\sout{at})uniroma1.it (G. Jona Lasinio).}

\begin{abstract}
%\textit{Shape} is all the geometrical information that remains when location, scale and rotational effects are removed from an object.
%Modelling shape data has recently gained interest in many research fields
%since any proper shape inference should rely on the geometric structure of the objects
%under study. Relevant applied fields are e.g.\ biology, medical imaging, computational anatomy, computer vision, archaeology and chemistry.
%From a theoretical point of view, the involved data spaces are not Euclidean hence differential geometric tools should be adopted in order to develop a proper statistical shape theory. Some descriptive and basic inference
%tools have been proposed in literature but to the current knowledge no regression methodologies have been proposed a part from \cite{dryden1},
%who propose a marginal likelihood approach to implement a size-and-shape
%response regression with Gaussian landmarks.
%Likelihood-based approaches do not efficiently work with complicated covariance structures and moreover could be compromised by convergence issues of numerical optimization algorithms. An alternative easily-interpretable
%approach which fits into the framework of Bayesian latent variable models
%potentially allows the specification of more complicated covariance structures
%and only requires a posterior approximation by means of MCMC methods.
Building on \cite{dryden1}, this note presents the Bayesian estimation of a regression model for size-and-shape response variables with Gaussian landmarks. Our proposal fits into the framework of Bayesian latent variable models and allows a highly flexible modelling framework.% which is still missing in the statistical literature. %\textit{Shape} is all the geometrical information that remains when location, scale and rotational effects are removed from an object.
\end{abstract}


\section{Introduction} 
Shape data naturally arise in various research fields, whenever our attention is on the shape of objects. For example, in biology there is  interest in modelling the shape features of organisms and understanding their relationship with environmental conditions. Even magnetic resonance data can be interpreted as containing size-and-shape information  \citep{mardia2013}. Image analysis, computer vision, and bioinformatics are other fields where shape data analysis finds natural applications \citep{anderson,greenmardia}. In genetics, electrophoretic gel images contain shape information, while chemistry employs three-dimensional coordinates to assess the geometric structure of molecules  \citep{horgan, czoigiel}.

Recent advances in this area focus on developing methodologies to capture and explain changes in the shape of objects. For example, \cite{kenobi} proposed smoothing methods, while, on the other hand, regression methodologies for shape and size-and-shape data have been explored in a limited number of papers. For instance, \cite{gutierrez2019} propose a Bayesian  approach to shape data, starting with a Gaussian distribution on the configuration space and then accounting for location, rotation, and scale effects using a projected normal distribution.
\cite{dryden1} tackle size-and-shape data, developing their proposal within the likelihood-based framework. An earlier paper \citep{dryden2} proposes the Bayesian estimation of a regression model where an additional parameter for the rotation matrix is introduced. However, this approach limitation is that it depends on defining a prior distribution on the rotation matrix, making the inference reliant on the rotation information of the original data. On the contrary, for size-and-shape modelling, the inference should be rotation-independent. Additionally,  landmarks are assumed independent.

This paper, along with \cite{dryden1},  treats the rotation as a latent variable, eliminating the need to specify a prior distribution. Additionally, the model considers dependent landmarks. As a result, this study presents an alternative Bayesian regression methodology tailored for size-and-shape response data. 
%This model is seamlessly integrated into the Bayesian latent variable models' framework.
The proposed methodology ensures estimation stability through a simple and efficient MCMC (Markov Chain Monte Carlo) algorithm. The model has been implemented in the Julia \citep{bezanson2017julia} package \emph{BayesSizeAndShape} \citep{jonasize}. 
%, demonstrating the commitment to offering accessible and practical tools for the research community.\\
%Utilizing this Bayesian regression methodology opens up exciting possibilities for further advancements in the analysis of size-and-shape data spanning various research fields. The study contributes to the ongoing evolution of shape data analysis and encourages more comprehensive investigations in the future.
 
\section{A Bayesian model for size-and-shape}

Let $\tilde{\mathbf{X}}_{i} \in \mathbb{R}^{(k+1)\times p}$, with $i=1,\dots, n$ and $k\geq p$, be a collection of random configuration matrices. In statistical shape analysis, the configuration matrix contains the Euclidean coordinates of $k+1$ landmarks 
(``a landmark is a point of correspondence on each object that matches between and within populations'',  \cite{drydenmardia2016})
%(i.e., reliably measured points on the object's boundary) 
in a $p$-dimensional space, where $p$ is typically 2 or 3. To perform size-and-shape inference, it is essential to eliminate information about location and orientation from every configuration matrix $\tilde{\mathbf{X}}_{i}$.
%Let $\tilde{\mathbf{X}}_{i} \in \mathbb{R}^{(k+1)\times p}$, with $i=1,\dots, n$,  be a $(k+1)\times p$ dimensional configuration matrix, with $k\geq p$, that represents the Euclidean coordinates of $k+1$ landmarks (i.e., points on the object's boundary that can be identified reliably) in dimension $p$ for the $i-$th recorded object, where $p$ is usually equal to 2 or 3. To perform any size-and-shape inference, we must remove information about the objects' location and orientation. 
To remove location information, we use the Helmert submatrix $\mathbf{H}$ \citep[see][]{drydenmardia2016}, where $\mathbf{H}$ has dimension  {$k\times (k+1)$}, and, setting $d_j = 1/\sqrt{j(j+1)}$, its $j$-th row
has the first $j$ elements  equal to $-d_j$, element $j+1$ is equal to  $jd_j$, and the remaining $k-j$ elements are equal to zero. By pre-multiplying the $i$-th configuration matrix with $\mathbf{H}$, we obtain the Helmertized configuration
%\begin{equation}
  $  \tilde{\mathbf{X}}_i^H := \mathbf{H}\tilde{\mathbf{X}}_{i}$.
%\end{equation}
%The matrix $\tilde{\mathbf{X}}_i^H \in \mathbb{R}^{k\times p}$ is commonly referred to as the \emph{pre-form} matrix. 
Following the approach in \cite{dryden1}, we can decompose each $\tilde{\mathbf{X}}_i^H$ using the singular value decomposition, which results in
%\begin{equation}
  $  \tilde{\mathbf{X}}_i^H = \mathbf{U}_i\boldsymbol{\Delta}_i\tilde{\mathbf{R}}_i^\top,$
%\end{equation}
where $\tilde{\mathbf{R}}_i \in \text{SO}(p)$, and $\text{SO}(p)$ represents the $p \times p$-dimensional Special Orthogonal group. The matrix $\tilde{\mathbf{R}}_i$ contains all the information about the $i$-th object's orientation. On the other hand, $\mathbf{Y}_i:=\mathbf{U}_i\boldsymbol{\Delta}_i$ represents the {size-and-shape} version of the original configuration $\tilde{\mathbf{X}}_i$, which is the actual focus of the inference.
By assuming $\tilde{\mathbf{R}}_i \in \text{SO}(p)$, we preserve reflection information that would otherwise be lost if $\tilde{\mathbf{R}}_i$ belongs to the space of the $p \times p$-dimensional Orthogonal group, denoted as $\text{O}(p)$. 
%This choice enhances the modelling capabilities and retains important orientation details.
%where we assume $\tilde{\mathbf{R}}_i \in \text{SO}(p)$, i.e., it belongs to the space of the $p \times p$-dimensional Special Orthogonal group, it could be readily proven that $\tilde{\mathbf{R}}_i$ contains all the information about orientation, $\mathbf{Y}_i:=\mathbf{U}_i\boldsymbol{\Delta}_i$  is the \textit{size-and-shape} version of the original configuration  $\tilde{\mathbf{X}}_i$, and it represents the object of the inference and the data we are modelling.  We assume $\tilde{\mathbf{R}}_i \in \text{SO}(p)$ to retain reflection information which is otherwise lost if $\tilde{\mathbf{R}}_i \in \text{O}(p)$.

\subsection{The model}
In our scenario, each $\mathbf{Y}_{i}$ is coupled with a vector of $d$ covariates, $\mathbf{z}_i = (z_{i1},z_{i2}, \dots , z_{id})^\top$, and we are interested in understanding the relationship between $\mathbf{z}_i$ and $\mathbf{Y}_{i}$. However, the size-and-shape space, where $\mathbf{Y}_{i}$ takes values, is a non-Euclidean manifold with a complex geometric structure, and directly specifying a probability distribution for $\mathbf{Y}_{i}$ is not a simple task.
We adopt the idea proposed in \cite{dryden1} to address these challenges, extending it to a Bayesian setting. The approach involves introducing a latent variable $\mathbf{R}_i \in \text{SO}(p)$ and defining a regressive-type relationship between $\mathbf{X}_{i}:=\mathbf{Y}_{i}\mathbf{R}_i^\top$ and $\mathbf{z}_i$ as follows:
\begin{equation}\label{eq:model1}
\text{vec}({\mathbf{X}_i}) \sim\mathcal{N}_{k p}\Big(  \text{vec}\Big(\sum_{h=1}^dz_{ih}\mathbf{B}_h\Big), \mathbf{I}_p \otimes \boldsymbol{\Sigma}\Big), \quad i = 1, \dots , n, 
\end{equation}
with  $\mathbf{X}_i \perp \mathbf{X}_{i'} $ if $i \neq i'$, and $\mathcal{N}_{kp}$ denotes the $kp$-dimensional Gaussian distribution. In the above expression, $\text{vec}(\cdot)$ denotes the vectorization of a matrix,  $\boldsymbol{\Sigma}$  is a non-singular $k \times k$ covariance matrix and $\mathbf{B}_h$, $h = 1, \dots , d$, is a $k\times p$ matrix of regressive coefficients. Remark that $\mathbf{R}_i$ is latent, and it must not be confused with $\tilde{\mathbf{R}}_i$, which is the rotation matrix of the original data. 
The latent variable approach avoids dependence on the rotation and leads to a proper inference depending only on $\mathbf{Y}_1, \dots, \mathbf{Y}_n$.
To describe our Bayesian model, we must introduce the prior distributions for the model's unknown parameters.   To simplify notation and the derivation of the full conditionals (i.e., the distribution of one parameter given all the others and the data) we use $\mathbf{X}_{i,l}$ and $\mathbf{B}_{h,l}$ to indicate the $l-$th column of $\mathbf{X}_i$ and $\mathbf{B}_h$, respectively, we assume %$\boldsymbol{\beta}_l = (\text{vec}(\mathbf{B}_{1,l}),\text{vec}%(\mathbf{B}_{2,l}),\dots, \text{vec}(\mathbf{B}_{d,l}))^\top$, 
$\boldsymbol{\beta}_l:=(\mathbf{B}_{1,l}^\top,\mathbf{B}_{2,l}^\top, \dots, \mathbf{B}_{d,l}^\top)^\top$,
$\boldsymbol{\beta}: = (\boldsymbol{\beta}_1^\top,\dots , \boldsymbol{\beta}_p^\top)^\top$, and we introduce the design matrix $\mathbf{Z}_{i} := \mathbf{I}_{k} \otimes \mathbf{z_i}^\top$. Since the $l$-th column of $\text{vec}(\sum_{h=1}^dz_{ih}\mathbf{B}_h)$ is equal to $\mathbf{Z}_{i} \boldsymbol{\beta}_l$, we define the Bayesian model as
\begin{align}
\mathbf{X}_{i,l}| \boldsymbol{\beta},\boldsymbol{\Sigma} &\sim\mathcal{N}_{k}(  \mathbf{Z}_{i} \boldsymbol{\beta}_l, \boldsymbol{\Sigma}), \quad i = 1, \dots , n,\quad l = 1,\dots,p, \label{eq:bay1}\\
\boldsymbol{\beta}_l & \sim \mathcal{N}_{kd}(\mathbf{M}_l, \mathbf{V}_l),\notag \\
\boldsymbol{\Sigma} & \sim \mathcal{IW}(\nu, \boldsymbol{\Psi}),\notag
\end{align}
with $\mathbf{X}_{i,l}|\boldsymbol{\beta},\boldsymbol{\Sigma} \perp \mathbf{X}_{i',l'}|\boldsymbol{\beta},\boldsymbol{\Sigma}$ if $i\neq i'$ or $l \neq l'$ and $\mathcal{IW}$ denotes the Inverse Wishart distribution.
It should be noted that, once we condition on $(\boldsymbol{\beta},\boldsymbol{\Sigma})$, models \eqref{eq:model1} and \eqref{eq:bay1} induce the same distribution on the configuration space. \\
Remark that an identification problem arises from the model specification, which has not been highlighted in \cite{dryden1}. To make it evident, we show that the sets of parameters $\{\mathbf{B}_1, \dots, \mathbf{B}_d, \boldsymbol{\Sigma}\}$ and $\{\mathbf{B}_1\boldsymbol{\Lambda}, \dots, \mathbf{B}_d\boldsymbol{\Lambda}, \boldsymbol{\Sigma}\}$, where $\boldsymbol{\Lambda} \in \text{SO}(p)$ is a rotation matrix, induce the same probability density function over $(\mathbf{U}_1, \Delta_1, \dots , \mathbf{U}_n, \Delta_n)$. Let  $\boldsymbol{\mu}_i: = \sum_{j=1}^dz_{ij}\mathbf{B}_j$ and $f$ denotes a probability density function, we need to prove that 
\begin{equation}\label{eq:rati}
    \frac{f(\mathbf{U}_i ,\Delta_i; \boldsymbol{\mu}_i, \boldsymbol{\Sigma})}{f(\mathbf{U}_i ,\Delta_i; \boldsymbol{\mu}_i\boldsymbol{\Lambda}, \boldsymbol{\Sigma})} =\exp\Big(-\frac{\text{tr}(\boldsymbol{\mu}_i^\top \boldsymbol{\Sigma}^{-1} \boldsymbol{\mu}_i) -  \text{tr}(\boldsymbol{\Lambda}^\top\boldsymbol{\mu}_i^\top\boldsymbol{\Sigma}^{-1} \boldsymbol{\mu}_i\boldsymbol{\Lambda})  }{2}\Big)= 1,
\end{equation}
for all $i=1, \dots , n$,  where the joint density of $(\mathbf{U}_i, \Delta_i)$ is derived by \cite{dryden1} in Theorem 1.
From the properties of the trace operator, and since $\boldsymbol{\Lambda}^\top = \boldsymbol{\Lambda}^{-1}$, we have that 
$
\text{tr}(\boldsymbol{\Lambda}^\top\boldsymbol{\mu}_i^\top\boldsymbol{\Sigma}^{-1} \boldsymbol{\mu}_i\boldsymbol{\Lambda}) = \text{tr}(\boldsymbol{\mu}_i^\top \boldsymbol{\Sigma}^{-1} \boldsymbol{\mu}_i),
$
which shows that \eqref{eq:rati} holds true for any $i=1,\dots,n$. For this reason, an identification constraint is needed to prevent any arbitrary rotation of $\boldsymbol{\mu}_i$. This can be achieved by assuming that for one of the $\mathbf{B}_h$, e.g., the first, we have that 
\begin{equation}\label{eq:cons}
[\mathbf{B}_1]_{wl} = 0, \quad l>w, \quad[\mathbf{B}_1]_{ll} \geq 0, \quad  l = 1,\dots, p-1.     
\end{equation}
These are a version of the transformations proposed by \cite{dryden1} to identify and isolate the size-and-shape information of the mean configuration.
%The transformation proposed by \cite{dryden1} aims to identify and isolate the size-and-shape information of the mean configuration $\boldsymbol{\mu}_i$. 
To impose these constraints effectively, two different approaches can be used.
The first approach involves modifying the prior distributions, altering the parameter space to adhere to the identification constraints. 
%The resulting posterior distribution will respect the desired properties by incorporating these constraints into the prior.
The second approach relies on the MCMC algorithm to explore the posterior freely, and after obtaining each posterior sample $\mathbf{B}_h^b$, where $b$ indicates the $b$-th sample, a remapping step is performed. This remapping transforms each $\mathbf{B}_h^b$ to an identified version $\tilde{\mathbf{B}}_h^b$ using the map
$\mathbf{B}_h^b\mapsto \tilde{\mathbf{B}}_h^b:=\mathbf{B}_h^b\boldsymbol{\Lambda}^{b}.$
Here, $\boldsymbol{\Lambda}^{b} = g(\mathbf{B}_h^b) \in \text{SO}(p)$ represents an appropriate rotation matrix defined by a function $g: \mathbb{R}^{k\times p}\to \text{SO}(p)$ such that  $\tilde{\mathbf{B}}_h^b$ satisfies the desired identification constraints  \eqref{eq:cons}.
We employ the second approach with a Gram-Schmidt construction to define $\boldsymbol{\Lambda}^b$. This method allows for a more straightforward MCMC algorithm, while respecting the constraints.

% These are a version of the transformation proposed by \cite{dryden1} to identify and isolate the size-and-shape information of the mean configuration $\boldsymbol{\mu}_i$.
% The constraints can be imposed in two different ways. The first one is by changing the prior distributions over the parameters that  change the parameter space to comply with the identification constraints. The other way is to let the Markov chain Monte Carlo (MCMC) algorithm explore freely the posterior and then remap  each $\mathbf{B}_h^b$, where $b$ indicates the $b$-th posterior sample, to an identified version $\tilde{\mathbf{B}}_h^b$, via the map 
% $\mathbf{B}_h^b\mapsto \tilde{\mathbf{B}}_h^b:=\mathbf{B}_h^b\boldsymbol{\Lambda}^{b},$
% where $\boldsymbol{\Lambda}^{b} = g(\mathbf{B}_h^b) \in \text{SO}(p)$ is an appropriate rotation matrix defined by a function $g: \mathbb{R}^{k\times p}\to \text{SO}(p)$, such that  $\tilde{\mathbf{B}}_h^b$ satisfies \eqref{eq:cons}. We use the latter, with a Gram–Schmidt construction to define $\boldsymbol{\Lambda}^b$, since it allows us to define a more straightforward MCMC algorithm.

\subsection{The Markov chain Monte Carlo algorithm}

To implement the MCMC algorithm, we need to derive the full conditional distributions of $\boldsymbol{\beta}$, $\boldsymbol{\Sigma}$, $\mathbf{R}_1$, $\mathbf{R}_2, \dots, \mathbf{R}_n$. Owing to the model specification given in  \eqref{eq:bay1}, we can easily see that the full conditional of $\boldsymbol{\beta}$ and $\boldsymbol{\Sigma}$ (indicated, respectively, as $\boldsymbol{\beta}_l|\cdots$ and $\boldsymbol{\Sigma}|\cdots$) are the same that we would obtain in the case of a standard Bayesian regression, i.e.,  
\begin{equation}
    \boldsymbol{\beta}_l|\dots  \sim \mathcal{N}_{kd}(\mathbf{M}_l^*, \mathbf{V}_l^*), \,\,\qquad
    \boldsymbol{\Sigma}|\dots  \sim \mathcal{IW}(\nu^*, \boldsymbol{\Psi}^*),
\end{equation}
with 
$$\mathbf{M}_l^*  =   \mathbf{V}_l^* \Big( \sum_{i=1}^n \mathbf{Z}_{i}^\top \boldsymbol{\Sigma}^{-1} \mathbf{X}_{i,l}  +  \mathbf{V}_l^{-1}  \mathbf{M}_l \Big), \quad  \mathbf{V}_l^*  = \Big( \sum_{i=1}^n \mathbf{Z}_{i}^\top\boldsymbol{\Sigma}^{-1} \mathbf{Z}_{i}  + \mathbf{V}_l^{-1} \Big)^{-1}$$
and 
$$  \nu^*  =  \nu + np, \quad \boldsymbol{\Psi}^*  = \boldsymbol{\Psi} + \sum_{i=1}^n \sum_{l=1}^p(\mathbf{X}_{i,l}-\mathbf{Z}_{i} \boldsymbol{\beta}_l)(\mathbf{X}_{i,l}-\mathbf{Z}_{i} \boldsymbol{\beta}_l)^\top.$$

To derive the full conditional distribution of $\mathbf{R}_i$, we can refer to the computation presented in \cite{dryden1}. According to Theorem 1, the distribution of $\mathbf{R}_i$ has density proportional to $\exp(\text{tr}(\mathbf{R}_i\mathbf{A}_i^\top))$ where $\mathbf{A}_i = \boldsymbol{\mu}_i^\top \boldsymbol{\Sigma}^{-1} \mathbf{Y}_i$, i.e. a Matrix Fisher distribution with parameter $\mathbf{A}_i$ \citep{mardiajupp2000}.
Sampling directly from a Matrix Fisher distribution is typically not straightforward, and ad hoc techniques are required; some proposals  can be found in \cite{kent} and \cite{Hoff2}. Here, we handle the case $p=2$ by expressing the rotation matrix as a function of the rotation angle $\theta_i \in [0, 2\pi)$. It can be shown that the distribution of $\theta_i$ is von-Mises, since its density is proportional to $\exp( \kappa_i \cos (\theta_i - \eta_i))$, and its parameters $(\eta_i,\kappa_i)$  can be  derived by setting the equation $\text{tr}(\mathbf{R}_i\mathbf{A}_i^\top) = \kappa_i \cos (\theta_i - \eta_i)$.
When $p=3$, we  express $\mathbf{R}_i$ as function of the Euler angles $\theta_{i,1}\in[0,2\pi)$, $\theta_{i,2}\in [0,\pi)$, $\theta_{i,3}\in[0,2\pi)$, by representing  $\mathbf{R}_i$ as the product of elementary rotations. Then, samples from the Matrix Fisher are obtained using the approach proposed by \cite{greenmardia}.

\section{Simulation experiments} 
The simulation study has been designed to assess the model's capacity to recover the underlying parameters used to generate the data. It covers a wide range of settings, providing a comprehensive evaluation of the model's performance under diverse conditions. We generate the data from equation \eqref{eq:bay1}, assuming that $\boldsymbol{\Sigma}$ is equal to $\lambda\boldsymbol{\Sigma}^*$ and $\lambda$ is a parameter we change in the simulations, with $d=3$. Variable ${z}_{i,1}=1$, $ \forall \, i=1,\dots,n$, defines the intercepts,  a continuous covariate $z_{i,2}$ is simulated from a Gaussian distribution with mean  10 and standard deviation  1, and a categorical variable $z_{i,3}$ with two levels is included in a corner-point representation. Each $z_{i,3}$ has an equal probability of assuming one of the two levels.
Parameter $\boldsymbol{\Sigma}$ is simulated from an  $\mathcal{IW}(k+2,5\mathbf{I}_k)$, and the regressive coefficients are generated from a  $\mathcal{N}(5,1)$. To comply with the identifiability constraint in \eqref{eq:cons}, coefficients are transformed to their identifiable version by multiplication of each $\mathbf{B}_{h}$ with an appropriate rotation matrix $\boldsymbol{\Lambda}$.
We generate 100 datasets for each of 16 different settings, considering all possible combinations of $k \in \{10,20\}$, $n \in \{20,100\}$, $\lambda\in\{1,10\}$, and $p \in\{2,3\}$. The total number of regressors to be estimated, taking into consideration the identifiability constraints,  are $kdp-1$ if $p=2$ and $kdp-3$ if $p=3$,   which ranges from a minimum of 59 to a maximum of 177, while the number of elements in $\boldsymbol{\Sigma}$ is $(k^2+k)/2$, which is equal to 55 if $k=10$ and 210 if $k=20$.
We assume $\mathbf{M}_l = \mathbf{0}_{kd}$, $\mathbf{V}_l = 10^{4}\mathbf{I}_{kd}$, $\nu = k+2$, and $\boldsymbol{\Psi} = \textbf{I}_k$. We run the algorithm for 90000 iterations, discarding the first 30000 iterations as burn-in, and keep every 30th iteration, retaining 2000 samples for inferential purposes. The model is implemented in Julia 1.8.2. %\citep{bezanson2017julia}, using the package \emph{BayesSizeAndShape}.

For each parameter, we evaluate the percentage of times that the 95\% credible interval (CI) contains the value used to simulate the data, and the total length of the CI. Tables \ref{tab:1} and \ref{tab:2} report the results of the simulation experiments. In particular, the proportion of correctly estimated parameters (parameters inside the associated CI) within the simulated datasets ranges from 0.917 to 0.974 under all settings. 
The CI lengths increase as $k$, $p$, and (especially) $\lambda$ increase. On the other hand, interval lengths decrease as $n$ increases, as expected.

\begin{table}[ht]
	\centering
	\caption{Simulation results with $p=3$. Columns $\%CI_{\beta}$ and $\%CI_{\Sigma}$ contain the percentage of  parameters, across the 100 simulated datasets, that are inside the associated 95\% CI.  Columns $L_{\beta}$ and  $L_{\Sigma}$ show the mean length of the 95\% CIs.}
	\medskip
	\begin{tabular}{lll|rrrr}
		\hline
		& &  &  $\%CI_{\beta}$ & $\%CI_{\Sigma}$ & $L_{\beta}$ & $L_{\Sigma}$ \\ 
		\hline
		$n=20$ & $k=10$ & $\lambda=1$ & 93.8& 91.9  & 3.02 & 2.73 \\ 
		$n=100$ & $k=10$ & $\lambda=1$ & 94.9 &93.6 & 1.25 & 1.09 \\ 
		$n=20$ & $k=20$ & $\lambda=1$  & 94.4& 94.2 & 3.17 & 3.96 \\ 
		$n=100$ & $k=20$ & $\lambda=1$  & 94.7& 93.3& 1.21 & 1.03 \\ 
		$n=20$ & $k=10$ & $\lambda=10$  & 94.2& 92.1 & 9.96 & 34.92 \\ 
		$n=100$ & $k=10$ & $\lambda=10$  & 94.1& 95.1  & 4.20 & 13.00 \\ 
		$n=20$ & $k=20$ & $\lambda=10$  & 94.0 &92.6& 11.60 & 53.32 \\ 
		$n=100$ & $k=20$ & $\lambda=10$  & 94.9& 94.0 & 3.78 & 9.44 \\ 
		\hline
	\end{tabular}  \label{tab:1}
\end{table}

\begin{table}[ht]
	\centering
	\caption{Simulation results with $p=2$. Columns $\%CI_{\beta}$ and  $\%CI_{\Sigma}$ contain the percentage of  parameters, across the 100 simulated datasets, that are inside the associated 95\% CI.  Columns $L_{\beta}$ and  $L_{\Sigma}$ show the mean length of the 95\% CIs.}
	\medskip
	\begin{tabular}{lll|rrrr}
		\hline
		& &  &  $\%CI_{\beta}$ & $\%CI_{\Sigma}$ & $L_{\beta}$ & $L_{\Sigma}$ \\
		\hline
		$n=20$ & $k=10$ & $\lambda=1$ &  96.7& 95.1 & 4.39 & 2.50 \\ 
		$n=100$ & $k=10$ & $\lambda=1$ & 94.8& 93.0 & 2.17 & 1.04 \\ 
		$n=20$ & $k=20$ & $\lambda=1$ &  95.5 &94.2 & 4.25 & 2.62 \\ 
		$n=100$ & $k=20$ & $\lambda=1$ &  94.6& 95.4 & 1.97 & 0.96 \\ 
		$n=20$ & $k=10$ & $\lambda=10$ &  95.9 &91.7 & 12.62 & 23.16 \\ 
		$n=100$ & $k=10$ & $\lambda=10$  & 97.1 &95.0 & 5.74 & 9.58 \\ 
		$n=20$ & $k=20$ & $\lambda=10$  & 97.4& 92.7 & 11.52 & 23.58 \\ 
		$n=100$ & $k=20$ & $\lambda=10$  & 96.5&94.9 & 5.42 & 9.62 \\ 
		\hline
	\end{tabular} \label{tab:2}
\end{table}

\section{Conclusions and further developments}
In this paper we presented the size-and-shape regression model within a Bayesian latent variable framework, addressing the identifiability issues.
Our work opens new possibilities for modelling size-and-shape data, particularly in formalizing complex dependence structures which might be handled within the Bayesian framework. Future research will explore dependence among landmarks using lattice-type modelling approaches.
We also aim to account for temporal dependence to model the shape evolution over time
%, enhancing understanding in various applications 
\citep{offsetnormal}.
Recognizing the limitations of linearity, we plan to explore more flexible functional relationships between covariates and mean configuration.
%In conclusion, our paper opens new avenues for size-and-shape data modeling, leveraging Bayesian latent variable techniques for addressing identifiability and facilitating exploration of diverse structures and relationships. 
%We anticipate further advancements in this field, enriching analysis in various research domains.

\section*{Acknowledgements}
The work of the second author was partially carried out within the FAIR - Future Artificial Intelligence Research and received funding from the European Union NextGenerationEU – Piano Nazionale di Ripresa e Resilienza (PNRR) – Missione 4, Componente 2, Investimento 1.3 – D.D. 1555 11/10/2022, PE00000013. This manuscript reflects only the author’s view and opinion, neither the European Union nor the European Commission can be considered responsible for them.

%This paper shows how an identifiability issue arises and can be solved in the Bayesian framework. 
%Finally, remark that a shape version of the model could be investigated by adding a latent variable representing the centroid size of the Helmertized configuration and standardizing by conditionally simulating from its size distribution. Clearly, a closed form for the size distribution could be cumbersome but simulating from it should be easily achieved once we get a sample from the Helmertized configuration. 

\bibliographystyle{apalike} 
\bibliography{bib}

\begin{thebibliography}{}

\bibitem[Anderson, 1997]{anderson}
Anderson, C.~R. (1997).
\newblock {\em Object recognition using statistical shape analysis}.
\newblock PhD thesis, University of Leeds.

\bibitem[Bezanson et~al., 2017]{bezanson2017julia}
Bezanson, J., Edelman, A., Karpinski, S., and Shah, V.~B. (2017).
\newblock {Julia: A Fresh Approach to Numerical Computing}.
\newblock {\em SIAM Review}, 59(1):65--98.

\bibitem[Czogiel et~al., 2011]{czoigiel}
Czogiel, I., Dryden, I.~L., and Brignell, C.~J. (2011).
\newblock {Bayesian matching of unlabeled marked point sets using random
  fields, with an application to molecular alignment}.
\newblock {\em Annals of Applied Statistics}, 5:2603--2629.

\bibitem[Dryden et~al., 2019]{dryden2}
Dryden, I.~L., Kim, K.-R., and Le, H. (2019).
\newblock {Bayesian Linear Size-and-Shape Regression with Applications to Face
  Data}.
\newblock {\em Sankhya A}, 81(1):83--103.

\bibitem[Dryden et~al., 2021]{dryden1}
Dryden, I.~L., Kume, A., Paine, P.~J., and Wood, A. T.~A. (2021).
\newblock {Regression Modeling for Size-and-Shape Data Based on a Gaussian
  Model for Landmarks}.
\newblock {\em Journal of the American Statistical Association},
  116(534):1011--1022.

\bibitem[Dryden and Mardia, 2016]{drydenmardia2016}
Dryden, I.~L. and Mardia, K. (2016).
\newblock {\em Statistical Shape Analysis: With Applications in R}.
\newblock Wiley Series in Probability and Statistics. John Wiley \& Sons,
  Chichester, 2nd edition.

\bibitem[Fontanella et~al., 2018]{offsetnormal}
Fontanella, L., Ippoliti, L., and Kume, A. (2018).
\newblock {The Offset Normal Shape Distribution for Dynamic Shape Analysis}.
\newblock {\em Journal of Computational and Graphical Statistics}.

\bibitem[Green and Mardia, 2006]{greenmardia}
Green, P.~J. and Mardia, K.~V. (2006).
\newblock {Bayesian alignment using hierarchical models, with applications in
  protein bioinformatics}.
\newblock {\em Biometrika}, 93:235--254.

\bibitem[Guti{\'e}rrez et~al., 2019]{gutierrez2019}
Guti{\'e}rrez, L., Guti{\'e}rrez-Pe{\~n}a, E., and Mena, R.~H. (2019).
\newblock {A Bayesian Approach to Statistical Shape Analysis via the Projected
  Normal Distribution}.
\newblock {\em Bayesian Analysis}, 14(2):427 -- 447.

\bibitem[Hoff, 2009]{Hoff2}
Hoff, P.~D. (2009).
\newblock {Simulation of the Matrix Bingham--von Mises--Fisher Distribution,
  With Applications to Multivariate and Relational Data}.
\newblock {\em Journal of Computational and Graphical Statistics},
  18(2):438--456.

\bibitem[Horgan et~al., 1992]{horgan}
Horgan, G.~W., Creasey, A., and Fenton, B. (1992).
\newblock {Superimposing two-dimensional gels to study genetic variation in
  malaria parasites}.
\newblock {\em Electrophoresis}, 13:871--875.

\bibitem[Kenobi et~al., 2010]{kenobi}
Kenobi, K., Dryden, I.~L., and Le, H. (2010).
\newblock {Shape Curves and Geodesic Modelling}.
\newblock {\em Biometrika}, 97(3):567--584.

\bibitem[Kent et~al., 2013]{kent}
Kent, J.~T., Ganeiber, A.~M., and Mardia, K.~V. (2013).
\newblock {A new method to simulate the Bingham and related distributions in
  directional data analysis with applications}.
\newblock {\em arXiv preprint arXiv:1310.8110v1}.

\bibitem[Mardia and Jupp, 2000]{mardiajupp2000}
Mardia, K. and Jupp, P. (2000).
\newblock {\em Directional Statistics}.
\newblock John Wiley \& Sons, New York, 1st edition.

\bibitem[Mardia et~al., 2013]{mardia2013}
Mardia, K.~V., Bookstein, F.~L., and Kent, J.~T. (2013).
\newblock {Alcohol, babies and the death penalty: Saving lives by analysing the
  shape of the brain}.
\newblock {\em Significance}, 10(3):12--16.

\bibitem[Mastrantonio and Jona~Lasinio, 2023]{jonasize}
Mastrantonio, G. and Jona~Lasinio, G. (2023).
\newblock {BayesSizeAndShape: a Julia package for Bayesian estimation of Size
  and Shape data regression models}.
\newblock In {\em Proceedings of the GRASPA 2023 Conference}.

\end{thebibliography}

\end{document}